\documentclass[referee,useAMS,usegraphicx,usenatbib]{mn2e}
\usepackage{times}

\newcommand{\Msun}{\ensuremath{\,{\rm M}_\odot}}                        
\newcommand{\Rsun}{\ensuremath{\,{\rm R}_\odot}}                        
\newcommand{\kms}{\,km\,s$^{-1}$}                                       

\title[T-Cyg1-12664: A low-mass chromospherically active eclipsing binary in the \textit{Kepler} field]
      {T-Cyg1-12664: A low-mass chromospherically active eclipsing binary in the \textit{Kepler} field}

\author[ \"{O}. \c{C}ak{\i}rl{\i} et al.]
  {\"{O}. \c{C}ak{\i}rl{\i}$^1$\thanks{e-mail:omur.cakirli@ege.edu.tr},  C.~\.{I}bano\v{g}lu$^1$, E.~Sipahi$^1$,\\ \\
  $^1$Ege University, Science Faculty, Astronomy and Space Sciences Dept., 35100 Bornova, \.{I}zmir, Turkey\\
  }

\begin{document} \maketitle 

\begin{abstract}
The eclipsing binary T-Cyg1-12664 was observed both spectroscopically and photometrically. Radial velocities of both components 
and ground-based VRI light curves were obtained. The Kepler's R-data and radial velocities for the system  
were analysed simultaneously. Masses and radii were obtained as 0.680$\pm$0.021 M$_{\odot}$ and 0.613$\pm$0.007 R$_{\odot}$for the primary and 
0.341$\pm$0.012M$_{\odot}$ and 0.897$\pm$0.012R$_{\odot}$ for the secondary star. The distance to the system was estimated as 
127$\pm$14 pc. The observed wave-like distortion at out-of-eclipse is modeled with two separate spots on the more massive star, which is 
also confirmed by the Ca {\sc ii}  K  and  H emission lines in its spectra. Locations of the components in the mass-radius and mass-effective 
temperature planes were compared with the well-determined eclipsing binaries' low-mass components as well as with the theoretical models. While 
the primary star's radius is consistent with the main-sequence stars, the radius of the less massive component appears to 
be 2.8 times larger than that of the main-sequence models. Comparison of the radii of low-mass stars with the models reveals that the 
observationally determined radii begin to deviate from the models with a mass of  0.27 \Msun~ and suddenly reaches to maximum deviation 
at a mass of 0.34 \Msun.~ Then, the deviations begin to decrease up to the solar mass. The maximum deviation seen at a mass of about 
0.34 \Msun~ is very close to the mass of fully convective stars as suggested by theoretical studies. A third star in the direction 
of the eclipsing pair has been detected from our VRI images. The observed infrared excess of the binary is most probably arisen 
from this star which may be radiated mostly in the infrared bands.   
\end{abstract}



\section{Introduction}                                                                                                               \label{sec:intro}
Transit events for the T-Cyg1-12664 were detected during regular operations of the \textit{Trans-Atlantic~Exoplanet~Survey} network 
({\em TrES}) for detecting the rare objects, such as transiting extrasolar planets and gravitational microlenses. \citet{2008AJ....135..850D} made 
identification and classification of 773 eclipsing binaries found in the {\em TrES}-survey. They have also estimated the absolute properties of 
these systems by a joint analysis of their light curves, colors and theoretical isochrones.  T-Cyg1-12664 is classified  
as a long-period single-lined eclipsing binary in their catalog, following a systematic analysis of the light curves within 10 fields of 
the {\em TrES} survey.  This binary was originally thought to have an 8.2-day orbital period because all of the eclipses observed in 
the {\em r}-passband of {\em TrES} have almost equal depths. This indicates that the system should be consisted of very similar components.  

Later on T-Cyg1-12664 was observed by the {\em Kepler} satellite in both long and short cadence and identified as KIC\,10935310 in 
the {\em Kepler} Input Catalog. A very shallow secondary eclipse is revealed by these observations. Detailed descriptions of the characteristics 
of these observations can be found in \cite{Sla}. The shallow secondary eclipse with respect to the primary minimum and the 
orbital period of about 4-day are indicative of the existence of a low mass star in the system. Therefore, we included T-Cyg1-12664 into our 
observing program on the low-mas stars. In this paper we present the results of analyses of the data gathered by {\em Kepler} 
and as well as our ground-based spectroscopic and photometric observations and {\em TrES}. We point out the special location of the components in the 
parameter space and offer to use it as a benchmark object for future theoretical studies of low-mass stellar objects.

\section{Data acquisition}                                                                                                             \label{sec:obs}
\subsection{Spectroscopy}
Optical spectroscopic observations of the T-Cyg1-12664 were obtained with the Turkish Faint Object Spectrograph Camera 
(TFOSC)\footnote{http://tug.tug.tubitak.gov.tr/rtt150\_tfosc.php} attached to the 1.5 m telescope in July, 2010, 
under good seeing conditions. Further details on the telescope and the spectrograph can be found at http://www.tug.tubitak.gov.tr. The 
wavelength coverage of each spectrum was 4000-9000 \AA~in 12 orders, with a resolving power of $\lambda$/$\Delta \lambda$ 
$\sim$7\,000 at 6563 \AA~and an average signal-to-noise ratio (S/N) was $\sim$120. We also obtained high S/N spectra of two M dwarfs  GJ\,740 
(M0\,V) and GJ\,182 (M0.5\,V) for use as templates in derivation of the radial velocities \citep{Nidever}. 

The electronic bias was removed from each image and we used the 'crreject' option for cosmic ray removal. Thus, the resulting 
spectra were largely cleaned from the cosmic rays. The echelle spectra were extracted and wavelength calibrated by using Fe-Ar 
lamp source with help of the IRAF {\sc echelle} package, see \citet{Tonry_Davis}.

\subsection{Photometric identification and follow-up observations}
T-Cyg1-12664 was first identified as a likely low-mass eclipsing binary candidate in the  \citet{2008AJ....135..850D}'s catalog, following a systematic analysis 
of the light curves (TrES; Alonso et al. 1996). The light curves of the survey consist of $\sim$5\,240 $Sloan$ $r$-band photometric measurements (depicted in 
Fig.\,7) binned to a 9-minute cadence. The calibration of the TrES images, the identification of stars and the extraction of the light curves are described by \citet{Dunham04}. 

T-Cyg1-12664 was observed as KIC\,10935310  between 24\,54\,953.0 - 24\,55\,370.7 by the $Kepler$ satellite. The short-cadence observations comprise 
135 and 170 data points obtained in the Quarters\,1 and 3, and the long-cadence data encompass 2\,651 and 2\,752 data points in the  Quarters\,0 
and 2, respectively. The uncorrected flux measurements are plotted against time in Fig.\,1. The 
times of the data are expressed in the barycentric Julian day (BJD) time-scale. The light curves of eclipsing binaries obtained by  $Kepler$ satellite 
have discontinuities with variable amplitudes due to one safe mode and four spacecraft attitude tweaks in Q2. Therefore, one encounters with some 
problems, especially variability in the baseline flux, when creating a complete light curve of a system. This may be occurred from either systematic 
effects (focus drifts, safe modes), intrinsic stellar variability (chromospheric activity, pulsations), or extrinsic contamination by a third light (a variable 
source that contributes light in the aperture of the object of interest). For these complications and other substantial data issues see the "{\it Kepler Data 
Characteristics Handbook}" \citep{chris}. The main Kepler pipeline conveys two kinds of photometric data: {\it calibrated} and {\it corrected}. Calibrated 
data are gained by performing pixel-level calibration that corrects for the bias, dark current, gain, nonlinearity, smear and flat field, and applies aperture 
photometry to reduced data. Corrected data are the result of detrended data which corrects degraded cadence due to data unorthodox and removes variability.

The original data of T-Cyg1-12664 and their detailed descriptions can be found in \cite{Sla}. Most notable for a chromospherically active star where 
spot modulation causes a significant baseline variability with an amplitude of the same order as the depth of the secondary eclipse in the light 
curves.  Fig.\,1 shows the uncorrected flux measurements plotted versus the BJD. These light curves clearly show variations on a time scale of 
about 4 days, due to the eclipse, superimposed on a small amplitude light fluctuations.  In Fig.2 (top panel) we plot the data against the BJD, excluding 
the eclipses. The light variations on a short time-scale have a peak-to-peak amplitude of about 0.020 in flux units. A second-order polynomial fit to the 
data at out-of-eclipse was obtained and subtracted from all of the fluxes, including eclipses. The remaining data after removal of the eclipses are 
plotted in the middle panel of Fig.2 which show again a variation with a cycle of about 90 days and a peak-to-peak amplitude of about 0.007 in 
flux units. The full data including eclipses are plotted in the bottom panel of Fig.2. The distortions on the light curve with changing amplitude 
are clearly seen.           

In order to provide ground-based observations of the system in different passbands, we used {\sf SI\,1100 CCD 
Camera\footnote{http://www.specinst.com/}} mounted on the 1\,m telescope at the T\"{U}B{\.I}TAK National Observatory, Turkey. 
Further details on the telescope and the instruments can be found in \cite{cakirli}. Differential aperture photometry was performed 
to obtain the light curves in the V-, R-, and I-bandpass. We iteratively selected comparison and check stars, namely 
GSC\,3562\,396 and GSC\,3561\,2134. The typical $rms$ residuals for each filter vary between 0.003 and 0.007 mag in differential 
magnitudes depending on the atmospheric conditions. The ground-based R-passband light curve is shown in the sixth panel from top-to-bottom of Fig.\,7.

\begin{figure}
\begin{center}
\includegraphics[width=11cm]{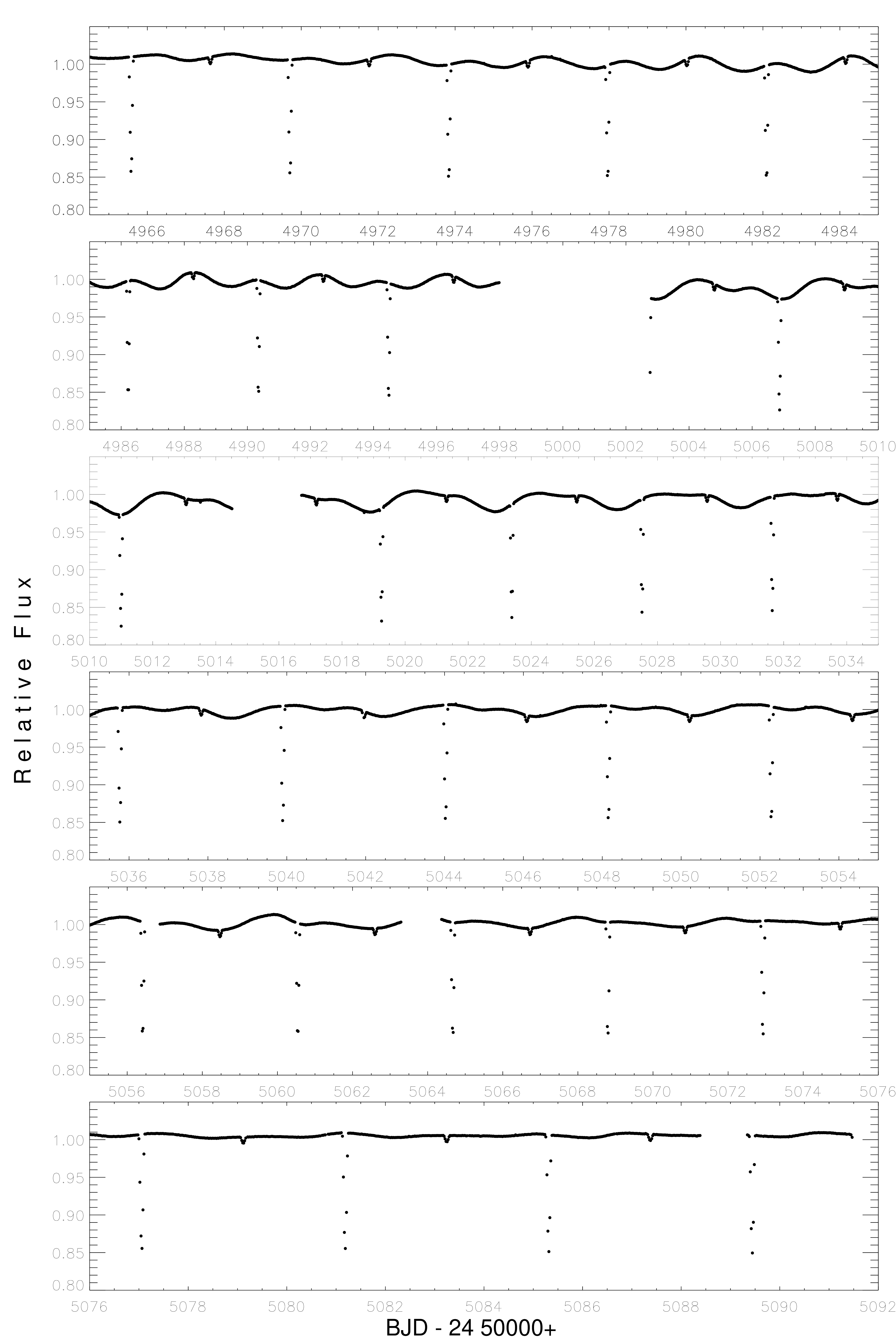}
\caption{Light curves of T-Cyg1-12664 obtained by Kepler.} \end{center} \end{figure}

\begin{figure*}
\begin{center}
\includegraphics[width=12cm]{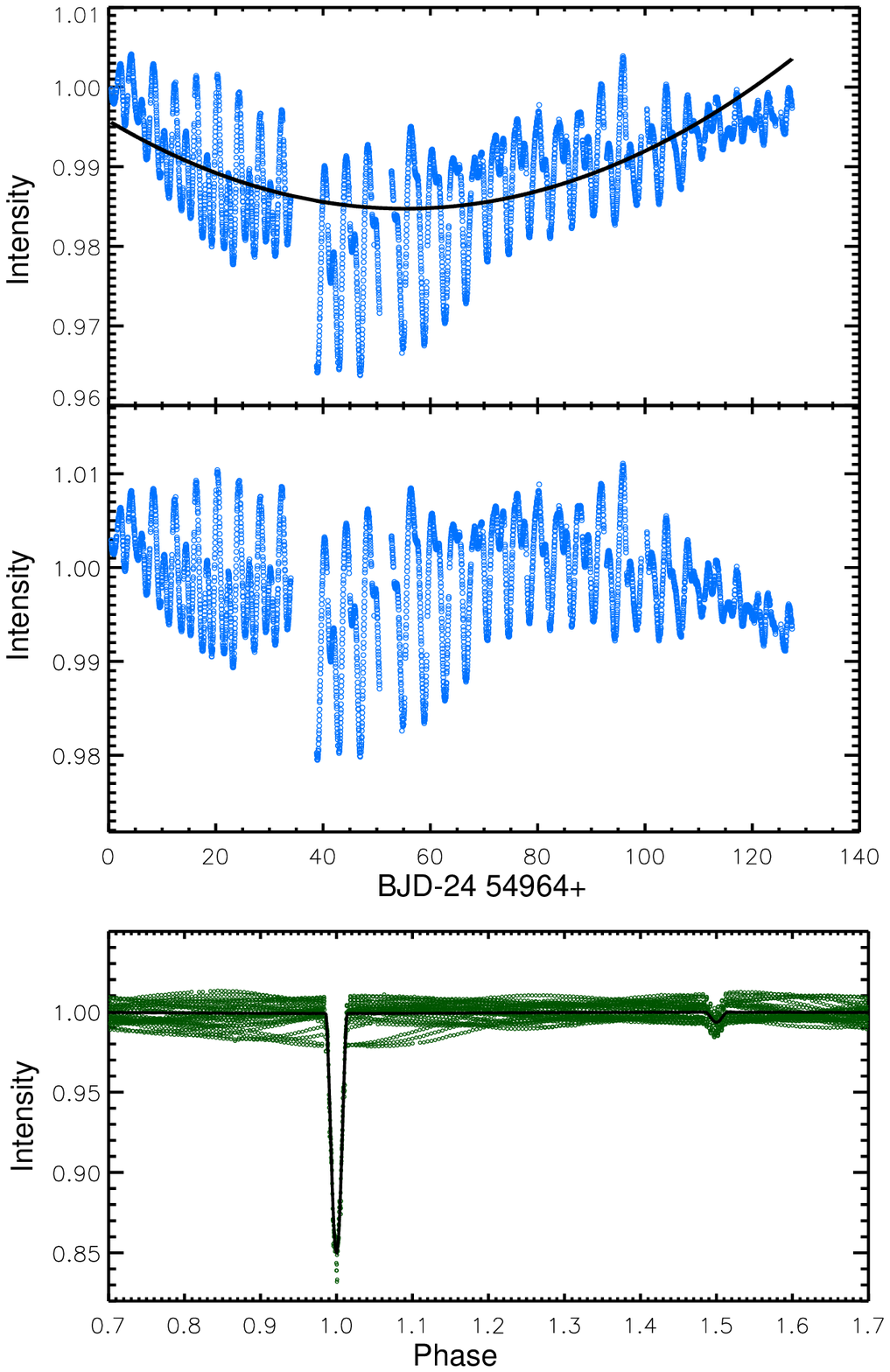}
\caption{Light curve of the T-Cyg1-12664 after extracting the eclipses (top panel) and its parabolic representation. Wave-like distortions on 
the out-of-eclipses are clearly visible. The remaining light curve after removal of the parabolic variation from the fluxes (middle 
panel). The final light curve including eclipses (bottom panel). In the top and middle panels the apsis are BJD while in the bottom 
panel is the orbital phase. The ordinates are fluxes.} \end{center} \end{figure*}

 \section{Analysis}
\subsection{Spectral classification}
We have used our spectra to reveal the spectral type of the primary component of T-Cyg1-12664. For this purpose we have degraded the 
spectral resolution from 7\,000 to 3\,000, by convolving them with a Gaussian kernel of the appropriate width, and we have measured
the equivalent widths ($EW$) of photospheric absorption lines for the spectral classification. We have followed the procedures of 
\citet{hernandez}, choosing helium lines in the blue-wavelength region, where the contribution of the secondary component to the 
observed spectrum is almost negligible. From several spectra we measured $EW_{\rm He I+ Fe I\lambda 4922 }=0.171\pm 0.088$\,\AA~and 
$EW_{\rm H_\beta 4861 }=3.24\pm 0.21$\,\AA. From the calibration relations $EW$--Spectral-type of \citet{hernandez}, we have derived 
a spectral type of K5 with an uncertainty of about 1 spectral subclass and an effective temperature of 4\,320$\pm$100 K from the 
tables of \citet{drill}.

The catalogs USNO, NOMAD and GSC2.3 provide (see Table\,1 for additional information) BVR$r$JHK magnitudes for T-Cyg1-12664 with 
a few tenths of a magnitude uncertainities. Using the USNO B=14.24$\pm$0.20 mag and GSC2.3 V=13.11$\pm$0.30 mag we obtained 
an observed color of B-V=1.13$\pm$0.36 mag. The color-spectral type relationship of \citet{drill} gives a color of B-V=1.15 mag for 
a K5-dwarf. Therefore, an interstellar reddening of E(B-V) is ignored.  The observed infrared colors of J-H=0.329$\pm$0.021 and 
H-K=0.053$\pm$0.021 are obtained using the JHK magnitudes given in the 2MASS catalog \citep{cutri}. These colors correspond to a 
main-sequence G4$\pm$2 star which is not consistent with that estimated from the spectra. This difference should be arisen from an 
additional star in the same direction with the eclipsing binary system. This star should radiate mostly at the infrared to change 
the color of the binary about one spectral class. Actually, a third star is seen in our CCD images which locates at the same 
direction with the binary system. A CCD image is shown in Fig.3.   

\begin{figure}
\includegraphics[width=12cm,angle=0]{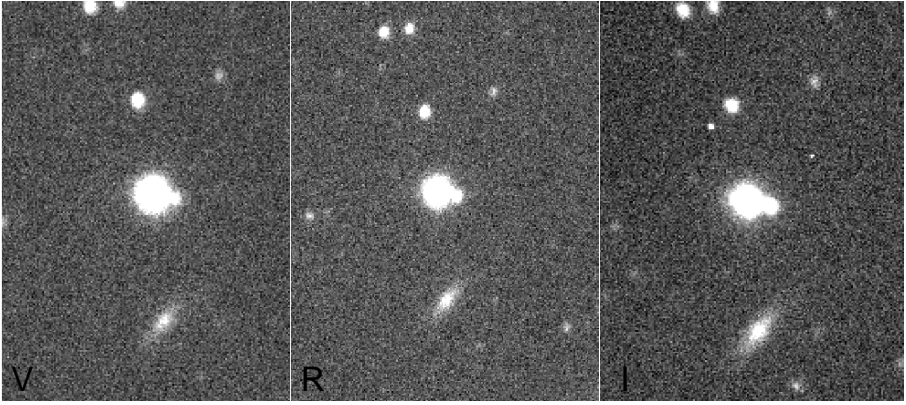}
\caption{\label{fig:periodresids} The 800x1500 pixels V, R, and I passbands  images of the field of binary system (Original size of the image 
is 20'x20'). A third star is clearly seen at the right side of the binary system (near the center). Note that the image of the third star is 
increasing in size towards the longer wavelengths. } \end{figure}

\begin{table}
\caption{Catalog information for T-Cyg1-12664.}
\begin{tabular}{lcc} 
\hline
Source Catalog & Parameter &Value\\
\hline
2MASS$^{a}$					    & $\alpha$ (J2000)      									& 19:51:39.8\\
2MASS 								& $\delta$ (J2000)      									& +48:19:55.4\\
USNO-B$^{b}$					& $B$ mag 													& 14.24 $\pm$ 0.20\\
GSC2.3$^{c}$					& $V$ mag 													& 13.11 $\pm$ 0.30\\
USNO-B 								& $R$ mag                  									& 13.32 $\pm$ 0.20\\			
CMC14$^{d}$						& $r'$ mag 													& 13.024 $\pm$ 0.035\\
2MASS  								& $J$ mag  													& 11.911 $\pm$ 0.015\\
2MASS 								& $H$ mag         											& 11.582 $\pm$ 0.015\\
2MASS 								& $K_s$ mag       										& 11.529 $\pm$ 0.015\\
USNO-B 								& $\mu_\alpha$ (${\rm mas\,yr^{-1}}$) 	&  $-$18 $\pm$ 6\\
USNO-B 								& $\mu_\delta$ (${\rm mas\,yr^{-1}}$) 	& $-$6 $\pm$ 2\\
$Kepler_{name}$				& KIC															& 10935310\\
$Kepler_{mag}$				    & K$_{mag}$												& 13.10$\pm$0.21\\
\hline \end{tabular} \\
\medskip
{\rm $^{a}${\em 2MASS} Two Micron All Sky Survey catalog \citep{Skrutskie06}.}, {\rm $^{b}${\em USNO-B} U.S. Naval Observatory photographic sky survey \citep{Monet03}.},
{\rm $^{c}${\em GSC2.3} Guide Star Catalog, version 2.3.2 \citep{Morrison01}.}, {\rm $^{d}${\em CMC14} Carlsberg Meridian Catalog 14 \citep{Evans02}.} 
\end{table}

\begin{table}
\caption{Times of minimum light for T-Cyg1-12664. The O-C values refer to the difference between the 
observed and calculated times of mid-eclipse.}
\label{O-C values}
\begin{tabular}{cccc}
\hline
  Minimum time  & Cycle number   & O-C & Ref.  \\
   (HJD-2\,400\,000)            & &     & \\
  \hline
 53177.8113	   &-431   &      0.0011  & 1	\\
 53181.9372	   &-430   &     -0.0018  & 1	\\
 53206.7121	   &-424   &      0.0003  & 1	\\
 53210.8415	   &-423    &      0.0009  & 1	\\
 53239.7411	   &-416    &     -0.0011  & 1	\\
 53243.8718	   &-415    &      0.0008  & 1	\\
 54965.5795(63)    &2     &     -0.0002  & 2	\\
 54969.7081(75)    &3     &     -0.0004  & 2	\\
 54973.8375(64)    &4     &      0.0003  & 2	\\
 54977.9656(10)    &5     &     -0.0005  & 2	\\
 54982.0941(28)    &6      &     -0.0007  & 2	\\
 54986.2240(31)    &7       &      0.0003  & 2	\\
 54990.3528(23)    &8       &      0.0004  & 2	\\
 54994.4814(15)    &9       &      0.0002  & 2	\\
 55006.8677(14)    &12      &      0.0001  & 2	\\
 55010.9963(13)    &13      &     -0.0002  & 2	\\
 55019.2535(25)    &15      &     -0.0006  & 2	\\
 55023.3823(30)    &16      &     -0.0005  & 2	\\
 55027.5126(41)    &17      &      0.0010  & 2	\\
 55031.6413(32)    &18      &      0.0009  & 2	\\
 55035.7699(22)    &1     &      0.0007  & 2	\\
 55039.8985(7)     &20     &      0.0004  & 2	\\
 55044.0262(47)    &21      &     -0.0007  & 2	\\
 55048.1556(8)     &22     &     -0.0000  & 2	\\
 55052.2847(61)    &23      &      0.0003  & 2	\\
 55060.5425(42)    &25      &      0.0005  & 2	\\
 55064.6712(29)    &26      &      0.0004  & 2	\\
 55068.8006(74)    &27      &      0.0010  & 2	\\
 55072.9283(8)     &28     &     -0.0001 & 2	\\
 55077.0566(6)     &29     &     -0.0007  & 2	\\
 55081.1849(61)    &30      &     -0.0012  & 2	\\
 55085.3142(9)     &31     &     -0.0007  & 2	\\
 55089.4433(4)     &32     &     -0.0003  & 2	\\
 55762.4377(4)     &195     &      0.0004  & 3	\\
\hline
\end{tabular}
\begin{list}{}{}
\item[Ref:]{\small (1) \citet{2008AJ....135..850D}, (2) $Kepler-data$, (3) VRI-data}
\end{list}
\end{table}

\subsection{Period determination}                                   
In the {\em TrES} observations the shallow secondary eclipse is not detected. Therefore, six times for the primary eclipse were obtained 
and the first ephemeris was presented by \citet{2008AJ....135..850D}. The secondary eclipse was clearly revealed by the  {\em Kepler} 
photometric observations. We determined 27 times for the mid-primary eclipse. In addition, we obtained one time for mid-primary eclipse 
from the ground-based photometric observations. Since the light curve is distorted, due to chromospheric activity, as large as the depth 
of the secondary eclipse, no attempt has been made for derivation of the times for the mid-secondary eclipse.  In order to measure the 
times of mid-primary eclipse we used the $\chi^2$  minimisation method. Very few of the minimum timings are accompanied by an 
errorbar, so uncertainties were assumed to be nearly equal. Since the times are given as BJD in the Kepler's data we transformed 
them into the heliocentric time scales. All times of minimum light of T-Cyg1-12664 were presented in Table 2. The orbital ephemeris 
given by \citet{2008AJ....135..850D} was used to determine the cycle numbers and the residuals between the observed and computed 
times of minima.  A straight line was fitted to the resulting cycle numbers and residuals, in the sense observed minus computed 
times, (Table\,2) by $\chi^2$ minimisation. The resulting  ephemeris is obtained as,

\begin{equation}
Min I(HJD)=2\,454\,957.3221(1)+4^d.12879779(64) \times E
\end{equation}
where the quantities in the parenthesis are the uncertainties in the final digits of the preceding number. The residuals of the 
fit are plotted in Fig.\,4. New orbital period is longer about 4 s than that found by \citet{2008AJ....135..850D}. Since the O-C 
residuals are represented by a straight line, although the observations gathered in three small time intervals but with large 
time spans between them, no substantive evidence is seen for departures from a constant period within six years. 

\begin{figure}
\includegraphics[width=12cm,angle=0]{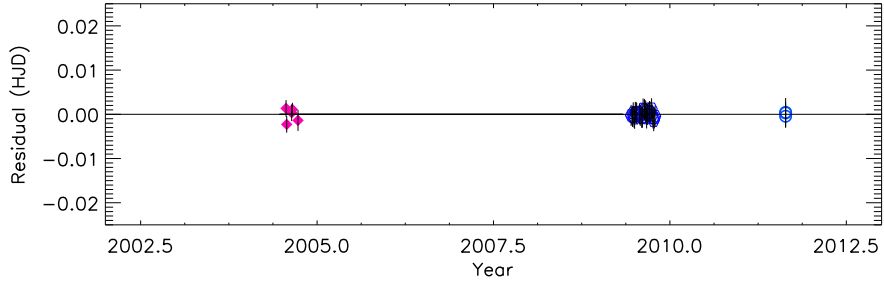}
\caption{\label{fig:periodresids} The residuals between the observed and computed (Eq.1) times of mid-eclipses.
} \end{figure}

\begin{figure}
\includegraphics[width=12cm]{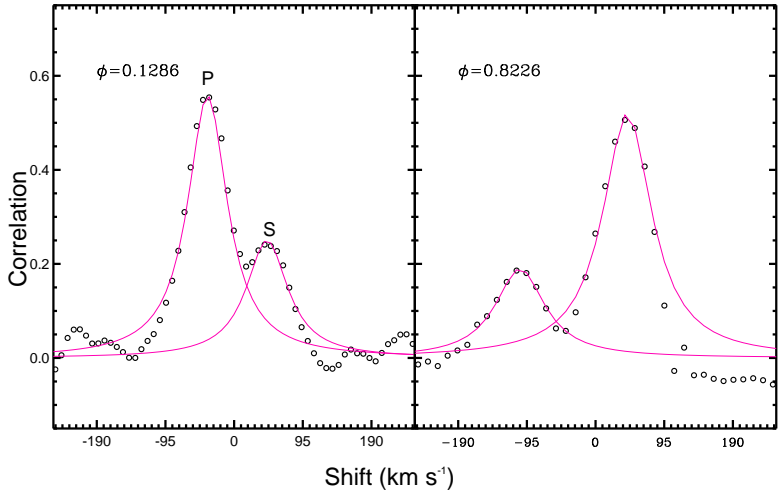}
\caption{\label{fig:CCF} Sample of CCFs between T-Cyg1-12664 and the radial velocity template spectrum around the first and second quadrature.}  
\end{figure}

\subsection{Radial velocity}
To derive the radial velocities of the components, the 9 TFOSC spectra of the eclipsing binary  were cross-correlated against 
the spectrum of GJ\,182, a single-lined M0.5\,V star, on an order-by-order basis using the {\sc fxcor} package in IRAF \citep{Simkin}. The majority of the spectra 
showed two distinct cross-correlation peaks in the quadrature, one for each component of the binary. Thus, both peaks were fitted independently
in the quadrature with a $Gaussian$ profile to measure the velocity and errors of the individual components. If the two peaks appear 
blended, a double Gaussian was applied to the combined profile using {\it de-blend} function in the task. For each of the 9 observations we 
then determined a weighted-average radial velocity for each star from all orders without significant contamination by telluric absorption
features. Here we used as weights the inverse of the variance of the radial velocity measurements in each order, as reported by {\sc fxcor}.

We adopted a $two-Gaussian$ fit algorithm to resolve cross-correlation peaks near the first and second quadratures when spectral lines are 
visible separately. Fig.\,5 shows examples of cros-correlations obtained by using the largest FWHM at nearly first and second quadratures. The 
two peaks correspond to each component of T-Cyg1-12664. The stronger peaks in each CCF correspond to the more luminous 
component which has a larger weight into the observed spectrum.

\begin{table}
\caption{Heliocentric radial velocities of T-Cyg1-12664. The columns give the heliocentric Julian date, the
orbital phase (according to the ephemeris in Eq.~1), the radial velocities of the two components with the 
corresponding standard deviations.}
\begin{tabular}{@{}ccccccccc@{}c}
\hline
HJD 2400000+ & Phase & \multicolumn{2}{c}{Star 1 }& \multicolumn{2}{c}{Star 2 } 	\\
             &       & $V_p$                      & $\sigma$                    & $V_s$   	& $\sigma$	\\
\hline
55390.47353 &	0.9098 &  22.3 &  4.8 &     --- &    --- \\
55391.36826 &	0.1265 & -35.8 &  3.2 &    58.9 &   5.3 \\
55392.52267 &	0.4061 & -28.2 &  4.6 &    44.1 &   7.7 \\
55393.39257 &	0.6168 &  25.5 &  6.6 &   -60.8 &   5.8 \\
55393.56632 &	0.6589 &  35.6 &  6.0 &   -82.5 &   4.8 \\
55394.40416 &	0.8618 &  32.3 &  4.2 &   -72.3 &   4.1 \\
55396.39116 &	0.3431 & -42.2 &  2.2 &    66.6 &   7.1 \\
55397.54163 &	0.6217 &  25.3 &  7.1 &   -67.7 &   8.5 \\
55398.36226 &	0.8205 &  37.6 &  3.3 &   -84.3 &   4.1 \\
\hline \\
\end{tabular}
\end{table}

The heliocentric radial velocities for the primary (V$_p$) and the secondary (V$_s$) components are listed in Table\,3 , along 
with the dates of observations and the corresponding orbital phases computed with the new ephemeris given in previous 
section. The velocities in this table have been corrected to the heliocentric reference system by adopting a radial velocity of 
14 \kms for the template star GJ\,182. The radial velocities listed in Table\,3 are the weighted averages of the values obtained 
from the cross-correlation of orders \#4, \#5, \#6 and \#7 of the target spectra with the corresponding order of the standard star 
spectrum. The weight $W_i = 1/\sigma_i^2$ has been given to each measurement. The standard errors of the weighted means have been 
calculated on the basis of the errors ($\sigma_i$) in the velocity values for each order according to the usual formula (e.g. 
Topping 1972). The $\sigma_i$ values are computed by {\sc fxcor} according to the fitted peak height, as described by \citet{Tonry_Davis}. The 
radial velocities are plotted against the orbital phase in Fig.\,6, which also includes the radial velocities measured by \citet{Devor}.

\begin{figure}
\includegraphics[width=10cm,angle=0]{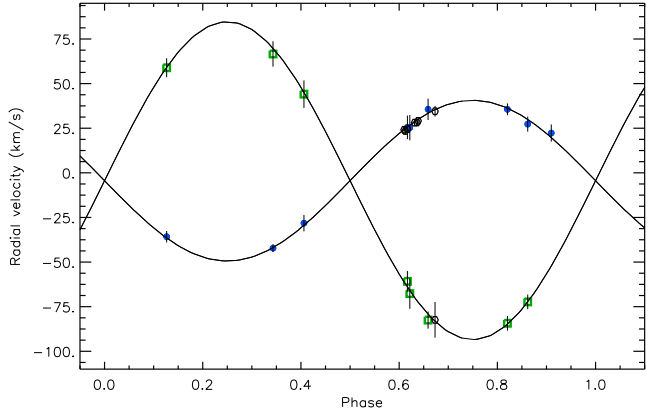}\\
\caption{Radial velocities folded on a period of 4.129-day and the model. Points with error bars (error bars are masked by 
the symbol size in some cases) show the radial velocity measurements for the components of the system 
(primary: filled circles, secondary: open squares). The velocities measured by \citet{Devor} are shown by empty circles. } \end{figure}

\subsection{Light curve and radial velocities modeling}
We used the most recent version of the eclipsing binary light curve modeling algorithm of \citet{wil}, as implemented in the 
{\sc phoebe} code of Pr{\v s}a \& Zwitter (2005), further developed by \citet{Wil} . In order to obtain the physical parameters of the component stars we, in 
the first step, analysed simultaneously  the detrended Kepler's R-data, plotted in the bottom panel of Fig. 2, and radial velocities 
shown in Fig. 6. The Wilson-Devinney code needs some input parameters, which depend upon the physical structures of the component 
stars. The values of these parameters can be estimated from global stellar properties. Therefore, we adopted the linear limb-darkening 
coefficients from Van Hamme (1993) as 0.635 and 0.602 for the primary and secondary components, respectively, taking into account the 
effective temperatures and the wavelengths of the observations; the bolometric albedos from \citet{lucy} as 0.5, typical for a 
convective stellar envelope, the gravity brightening coefficients as 0.32 for both components. The rotational velocities of the 
components are assumed to be synchronous with the orbital one.

The adjustable parameters in the light and RV curves fitting were the orbital 
inclination, the surface potentials of the two stars, the effective temperature of the secondary, and the color-dependent luminosity 
of the hotter star, the zero-epoch offset, semi-major axis of the orbit, the mass-ratio and the systemic velocity. A detached configuration 
(Mode 2) with coupling between luminosity and temperature was used for solution. The iterations were carried out automatically until 
convergence, and a solution was defined as the set of parameters for which the differential corrections were smaller than the probable 
errors. The computed light curve corresponding to the simultaneous light-velocity solution is compared with the observations in the 
bottom panel of Fig.2 and in Fig.6. 

It is obvious from these figures that the observed light curves and radial velocities are reproduced by the best fit theoretical curves shown by solid lines.
However, systematic deviations are apparent in the light curve, especially at out-of-eclipses. The light curve of the system is asymmetric in shape 
and varies with time. The wave-like distortion at out-of-eclipses in the light curves of T-Cyg1-12664 is attributed to the spot or spot groups on the 
primary star. The amplitudes of these variations exceed sometime the depth of the secondary eclipse. Therefore, we assumed that solar-like 
activity originates in the K5 star, maculation effects are separable from proximity and eclipse effects, and a cool, circular spot model may represent 
the parameters of the spotted regions. Due to the very small light contribution of the secondary star, as it is indicated by the preliminary 
analysis, the K5 star should be responsible for the magnetic activity which will be discussed in Sec.4. The second step was to include spot 
parameters as adjustable parameters. Using a trial-error method we estimated preliminary parameters for the spots. The shape of the 
distortion on the light curve depends upon the number of spots, locations and sizes on the star's surface. The trial-and-error method indicated 
that distortions could be represented by two cool spots on the primary star. Since the amplitude and shape of the wave-like distortion vary 
with time we divided the data into ten separate parts with a three-day intervals. The parameters calculated in the first step were taken as 
input parameters as well as spot parameters. The analysis was repeated and the final parameters and spot parameters are  listed in 
Tables\,4 and 5, respectively. The uncertainties assigned to the adjusted parameters are the internal errors provided directly by the 
Wilson-Devinney code. 

We also searched for the light contribution of the third star. Third light is added to the adjustable parameters but this trial is failed. This is because 
either light contribution of the third star is as small as to be ignored at the R-bandpass or its effect is overcome by the spot effects.                     

The computed light curves (continuous lines) are compared with the observations in Fig.\,7 (five panels from top to bottom) . In the bottom 
two panels of Fig.\,7  we also compare the computed curves with our R-data and the {\em TrES} r-passband data. The last two figures are 
presented only for comparison the observations with the computed curves obtained by the $Kepler$ data.

\begin{figure}
\includegraphics[width=10cm]{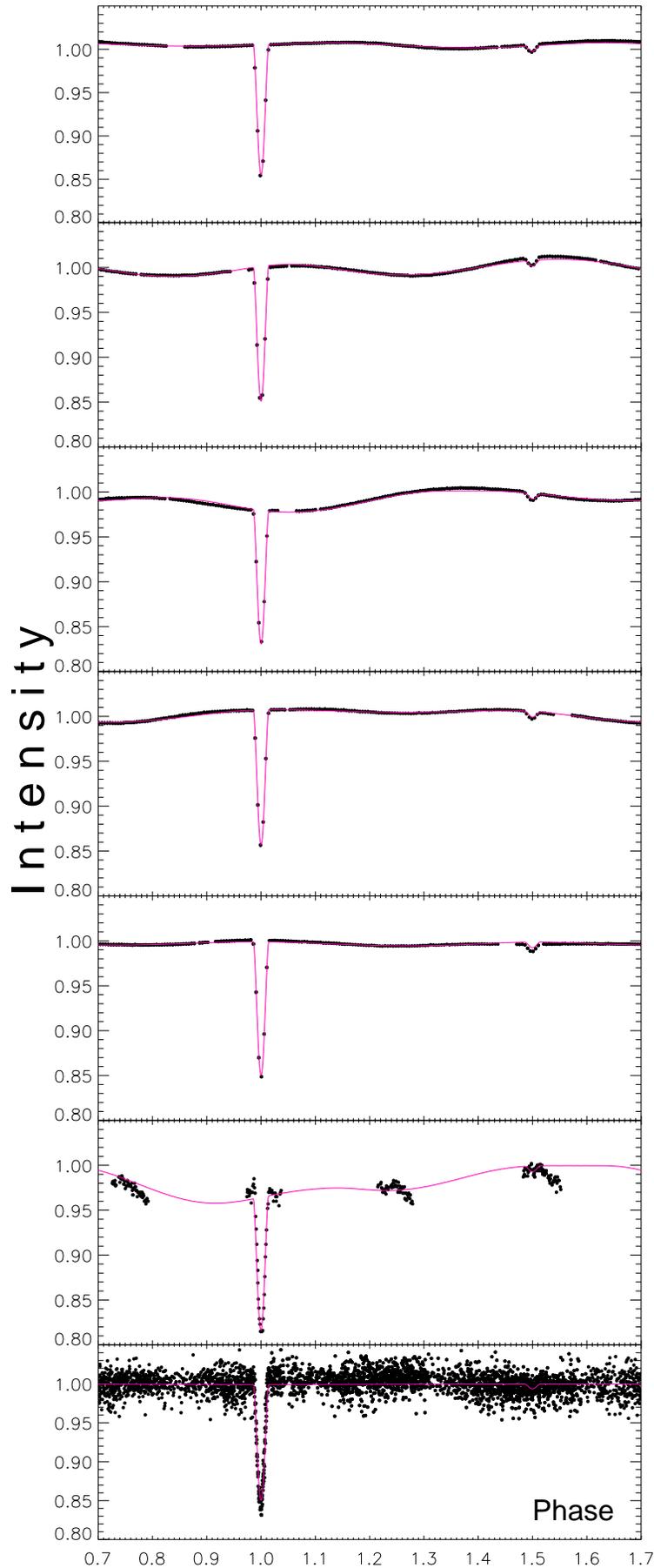}
\caption{\label{fig:periodresids} Sample light curves for T-Cyg1-12664 obtained in the $Kepler-R$-passband and the computed curves including spot parameters. In the sixth panel from top-to-bottom our ground-based R-passband light curve with spots and in the bottom panel the observations obtained by $TrES$ is compared with the computed light curve without spots. } \end{figure}

\begin{table}
\scriptsize
\caption{Results of the $Kepler-R$-passband light curves and radial velocities analyses for T-Cyg1-12664. }
\setlength{\tabcolsep}{0.8pt} 
\begin{tabular}{lc}
\hline
Parameters  &  $Kepler-R $  \\
\hline	
$i^{o}$																&83.84$\pm$0.04									\\
$\Omega_1$												&18.398$\pm$0.076					\\
$\Omega_2$												&7.341$\pm$0.049						\\
T$_{eff_1}$ (K)											&4320[Fix]											\\
T$_{eff_2}$ (K)											&2750$\pm$8									\\
$a$ (R$_{\odot}$)									    &10.904$\pm$0.016				\\
$V_{\gamma}$ (km s$^{-1}$)				                &-4.764$\pm$0.001 							\\   
$q$														&0.5021$\pm$0.0018	\\
r$_1$				   									&0.0562$\pm$0.0002					\\
r$_2$												    &0.0823$\pm$0.0007					\\
$\frac{L_{1}}{(L_{1}+L_{2})}$ 	       					&0.8980$\pm$0.0018				\\
$\chi^2$												&0.015						\\		
\hline
\end{tabular}
\end{table}

\begin{table*}
\centering
\footnotesize
\begin{minipage} {12 cm}
\caption{The spot parameters.}
\begin{tabular}{|l|c|c| c| c| c|}
\hline
\hline
BJD	& Spot 	&$Co-Lat$ ($^{\circ}$)    & $Co-Long$ ($^{\circ}$)	&	$R$ ($^{\circ}$) &  $T_{Spot}$/$T_{Sur}$ 	\\
\hline					
54965-54968 &  I  &	 51.4$\pm$2.5  &  49.1$\pm$1.2 & 12.4$\pm$0.3 & 0.9782$\pm$0.0012 \\
            &  II &	 93.4$\pm$1.1  & 220.2$\pm$0.8 & 20.2$\pm$0.4 & 0.9900$\pm$0.0003 \\
								
54982-54985 &  I  &	 14.9$\pm$1.1  &  59.9$\pm$0.3 & 17.3$\pm$0.1 & 0.9610$\pm$0.0006 \\
            &  II &      78.0$\pm$0.6  & 265.5$\pm$0.6 & 16.6$\pm$0.1 & 0.9621$\pm$0.0006 \\
								
55003-55006 &  I  &	106.8$\pm$1.3  & 115.0$\pm$1.5 & 15.7$\pm$0.2 & 0.9702$\pm$0.0008 \\
            &  II &	 91.6$\pm$0.4  & 340.0$\pm$0.7 & 19.9$\pm$0.1 & 0.9631$\pm$0.0005 \\
								
55034-55037 &  I  &	 45.2$\pm$4.1  & 100.2$\pm$1.2 & 18.3$\pm$0.2 & 0.9573$\pm$0.0005 \\
            &  II &	 85.7$\pm$25.7 & 260.5$\pm$0.6 &  9.9$\pm$0.1 & 0.9783$\pm$0.0010 \\
								
55079-55082 &  I  &	 97.7$\pm$0.9  &  99.9$\pm$1.3 & 15.2$\pm$0.3 & 0.9880$\pm$0.0006 \\
            &  II &	 48.8$\pm$1.3  & 260.0$\pm$1.6 & 16.1$\pm$0.3 & 0.9849$\pm$0.0006 \\
								
55760-55763 &  I  &	 90.7$\pm$10.7 &  30.8$\pm$1.9 & 16.1$\pm$0.2 & 0.8800$\pm$0.0042 \\
            &  II &	135.7$\pm$1.1  & 277.0$\pm$4.4 & 18.3$\pm$0.4 & 0.8800$\pm$0.0066 \\
\hline
\end{tabular}
\end{minipage}
\end{table*}
\smallskip

\subsection{Absolute parameters}
The orbital inclination and mean fractional radii of the components are found to be $i$=83.84$\pm$0.04, 
r$_1$=0.0562$\pm$0.0002, and r$_2$=0.0823$\pm$0.0007 from the light and radial velocity curves analyses. The 
separation between the components was estimated to be $a$=10.904$\pm$0.016 R$_{\odot}$. The effective temperature of 5770 K and bolometric
magnitude of 4.74 mag are adopted for the Sun. The standard deviations of the parameters have been determined by JKTABSDIM\footnote{This 
can be obtained from http://http://www.astro.keele.ac.uk/$\sim$jkt/codes.html} code, which calculates distance and 
other physical parameters using several different sources of bolometric corrections (Southworth et al. 2005). The best 
fitting parameters are listed in Table\,6 together with their formal standard deviations. The masses of the primary and 
secondary stars are determined with an accuracy of about 3\%  and 3.5\% , respectively, while the radii with 1 \% . We 
compared the position of the primary star in the age-radius diagram with the evolutionary tracks of \citet{gir00} for 
solar metallicity. The location of the primary star in the Hertzsprung-Russell diagram is in a good agreement with those 
of K5 main-sequence stars. We roughly estimate its age as 3.4 Gyr. This indicates that the secondary star should also 
be on the main-sequence band of the Hertzsprung-Russell diagram.     

The light contribution of the primary star is found to be 0.898 for the R-passband. Its apparent visual magnitude is calculated 
as 13.14$\pm$0.30 mag. The interstellar reddening of E(B-V)=0.0 mag yield a distance to the system as 126$\pm$13  pc for 
the bolometric correction of -0.72 mag for a K5 main-sequence star \citep{drill}. However, the mean distance to the system 
is estimated as 128$\pm$14 pc for the the bolometric corrections taken from  \citet{gir02}. The mean distance to the system 
was determined as 127$\pm$14 pc.

\subsection{Kinematics}
To study the kinematical properties of T-Cyg1-12664, we used the system's centre-of-mass velocity, distance and proper motion values. The 
proper motion data were taken from 2MASS catalogue \citep{Skrutskie06}, whereas the centre-of-mass velocity and 
distance are obtained in this study. The system's space velocity was calculated using \citet{sod87} algorithm. The U, V and W space 
velocity components and their errors were obtained and given in Table\,6. To obtain the space velocity precisely, the first-order galactic 
differential rotation correction was taken into account \citep{mih81}, and -1.08 and 0.65 \kms differential corrections were applied to 
U and V space velocity components, respectively. The W velocity is not affected in this first-order approximation. As for the LSR (Local Standard of Rest) correction, 
\citet{mih81} values (9, 12, 7)$_{\odot}$ \kms were used and the final space velocity of T-Cyg1-12664 was obtained as  {\it S}\,=17 \kms. This 
value is in agreement with space velocities of the young stars \citep{montes2001}.

\begin{table}
\setlength{\tabcolsep}{2.5pt} 
\caption{Fundamental parameters of T-Cyg1-12664.}
\label{parameters}
\begin{tabular}{lcc}
\hline
& \multicolumn{2}{c}{\hspace{0.25cm}T-Cyg1-12664} 																									\\
Parameter 												& Primary	&	Secondary																						\\
\hline
Mass (M$_{\odot}$) 								& 0.680$\pm$0.021 & 0.341$\pm$0.012																\\
Radius (R$_{\odot}$) 								& 0.613$\pm$0.007 & 0.897$\pm$0.012																\\
$T_{eff}$ (K)											& 4\,320$\pm$100	& 2\,750$\pm$65      																\\
$\log~g$ ($cgs$) 										& 4.696$\pm$0.006 & 4.065$\pm$0.011															\\
$\log~(L/L_{\odot})$								& -0.928$\pm$0.041	& -1.381$\pm$0.043       															\\
$(vsin~i)_{calc.}$ (km s$^{-1}$)			&7.51$\pm$0.08  &10.99$\pm$0.15       																		\\   
Spectral Type											& K5V$\pm$1  	&M3V$\pm$1  																		\\
$d$ (pc)													& \multicolumn{2}{c}{127$\pm$14}																	\\
$\mu_\alpha cos\delta$, $\mu_\delta$(mas yr$^{-1}$) & \multicolumn{2}{c}{-18$\pm$6, -6$\pm$2} 							\\
$U, V, W$ (km s$^{-1}$)  						& \multicolumn{2}{c}{4$\pm$1, 2$\pm$1, -7$\pm$1}										\\ 
\hline  
 \end{tabular}
\end{table}

\begin{figure}
\includegraphics[width=12cm]{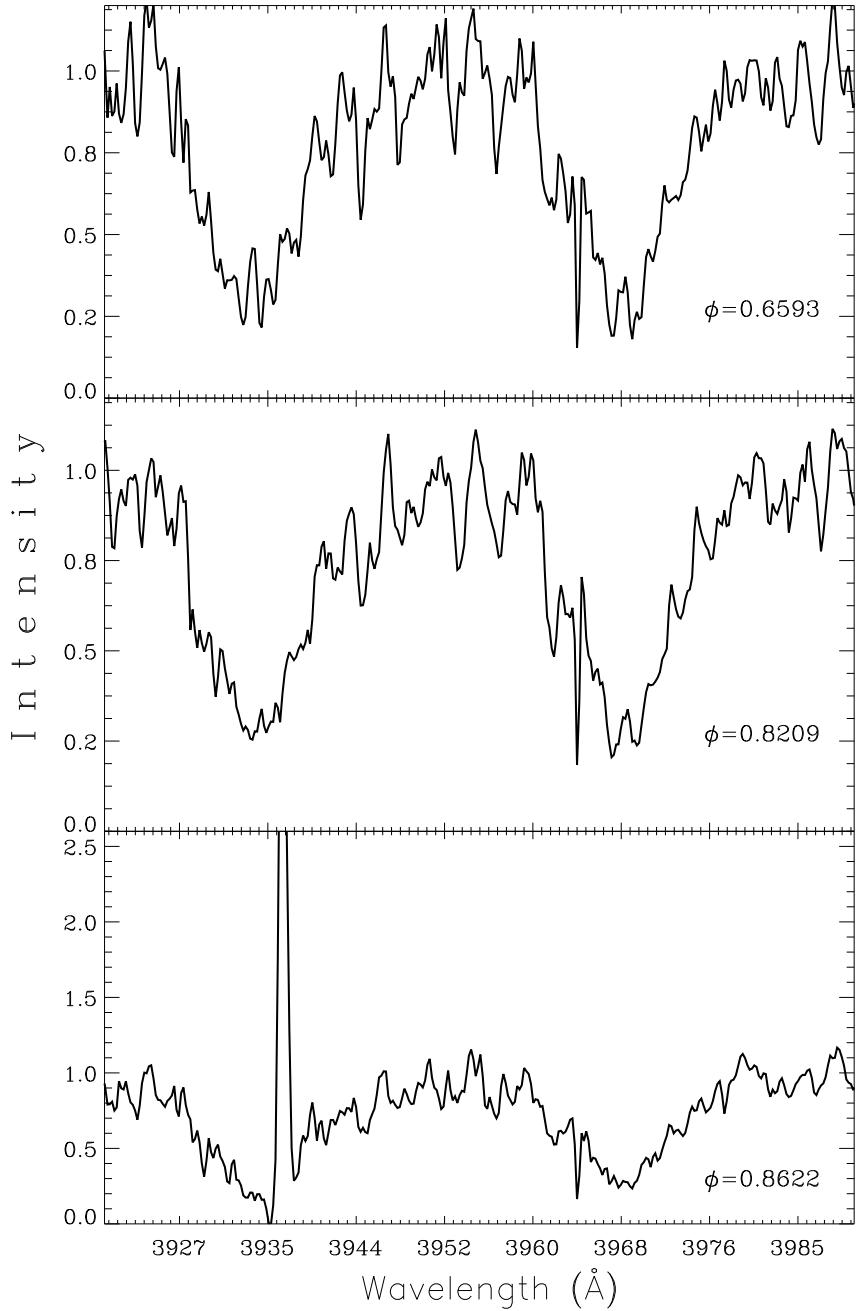}
\caption{Ca {\sc ii}-K (3933.68 \AA)\,and Ca {\sc ii}-H (3968.49 \AA) emission lines in the spectrum of the primary star.} \end{figure}

\section{Stellar activity}
The VRI light curves of T-Cyg1-12664 clearly show wave-like distortions, especially at out-of-eclipse. Such a cyclic variation is 
common properties of the active binary systems, i.e. RS CVn-type binaries. The amplitude of distortions on the light curves 
appears to change during the time span of the observations. The minimum and maximum amplitudes are measured as 5.9 
and 30.5 mmag, respectively. However, the amplitudes of the wave-like distortion in the ground-based observations are 
0.061, 0.042 and 0.033 mag in V-, R- and I-passband light curves, respectively. The effect of distortion is diminishing toward 
the longer wavelengths as it is expected for a K5 star. As it is evidenced from the top panel of Fig.2 the amplitude of the 
distortion increases up to about BJD 2455014 and then decreases again. A preliminary cycle for distortion is estimated to 
be about 127 days. The wave-like distortion has been represented by two spots located on the K5 star. The spot parameters 
are already given in Table 5. They are located at latitudes between 15 and 135 degrees, mostly in 40-100 degrees. It seems 
that they are located at longitudes of 90 and 270 degrees separated about 180 degrees. This result indicates that there are 
two active regions on the primary star. Such active longitude belts are common in chromospherically active stars 
\citep{Eat}, \citep{Uch}.  The amplitude of the distortion is increased when the spots are located at latitudes about 90 
and 107 degrees. The spots appears to be cooler about 140 K than the effective temperature of the primary star. In 
addition, the mean brightness level at the maxima, i.e. (MaxI+MaxII)/2, of the system varies of about 8 mmag with 
a cycle of about 90 days. The mean brightness has reached two local maxima and minima (see Fig. 2 middle panel).       

The mostly used indicator of the stellar activity is Ca {\sc ii} emission lines. In Fig.\,8 the Ca {\sc ii} - K (3933.68 Å) 
and Ca {\sc ii} - H (3968.49 Å) emission lines for primary are shown. In the spectrum of the system hydrogen absorption 
lines are usually shallow which is another indicator of the stellar activity. 

\begin{figure}
\includegraphics[width=12cm]{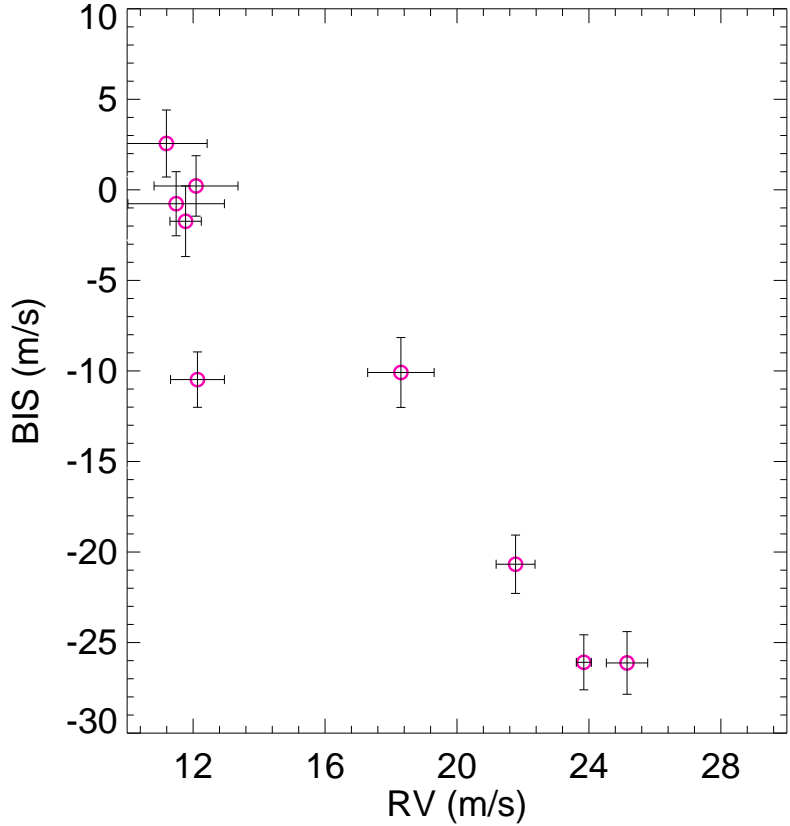}
\caption{Bisector velocity span vs. radial velocity for the system. The clear negative correlation indicates that radial velocity variations are due to stellar activity. } \end{figure}

We also carried out a line bisector (BIS) analysis to enable us to find out whether the light curves and line profile variations may be attributed to
starspots. It is a method describing quantitatively the tiny asymmetry or subtle changes in the line profiles. It is easily done by marking
the middle of horizontal cuts of profile in the different line depths. The line connecting such points (called bisector) is then zoomed in 
horizontal (wavelength or radial velocity) direction. The characteristic shapes of bisector are reported in the literature (see for example
\citep{Que98}; \citep{fio2005}); {\it i}) anti-correlation, which indicates that the radial velocity variations are due to stellar activity (by active 
regions at the stellar surface like spots or plages), {\it ii}) lack of correlation, which indicates the Doppler reflex motion around the centre-of-
mass due to the other bodies orbiting the star, {\it iii}) correlation, which, as pointed out by \citet{fio2005}, indicates that the radial velocity 
variations are due to light contamination from an unseen stellar companion. 

As shown in Fig.\,9, there is an anti-correlation between BIS and radial velocity, with a reliable correlation coefficient. This result suggests that the
radial velocity variations of T-Cyg1-12664  are due to stellar activity variations (e.g. spots on photosphere). 

\begin{table*}
\scriptsize
\setlength{\tabcolsep}{2.5pt} 
\caption{The mass, radius, effective temperature and orbital period of the low-mass stars in the double-lined eclipsing binaries. }
\begin{tabular}{lcrclrclrclccr}
\hline
Object	&Star	&\multicolumn{3}{c}{Mass/\Msun}	&\multicolumn{3}{c}{Radius/\Rsun}&\multicolumn{3}{c}{T$_{\rm eff}$/K}&P$_{\rm orb}/\rm day$	&Refs	&\\
\hline
CM Dra		&A	&0.2310	&$\pm$	&0.0009	&0.2534	&$\pm$	&0.0019	&3130	&$\pm$	&70	&1.276	&1	&\\
				&B	&0.2141	&$\pm$	&0.0008	&0.2398	&$\pm$	&0.0018	&3120	&$\pm$	&70	&	&	&\\
YY Gem	&A	&0.5992	&$\pm$	&0.0047	&0.6194	&$\pm$	&0.0057	&3820	&$\pm$	&100	&0.820	&2	&\\
				&B	&0.5992	&$\pm$	&0.0047	&0.6194	&$\pm$	&0.0057	&3820	&$\pm$	&100	&	&	&\\
CU Cnc		&A	&0.4349	&$\pm$	&0.0012	&0.4323	&$\pm$	&0.0055	&3160	&$\pm$	&150	&2.794	&3	&\\
				&B		&0.3992	&$\pm$	&0.0009	&0.3916	&$\pm$	&0.0094	&3125	&$\pm$	&150	&	&	&\\
GU Boo	&A	&0.6101	&$\pm$	&0.0064	&0.6270	&$\pm$	&0.0160	&3920	&$\pm$	&130	&0.492	&4	&\\
				&B	&0.5995	&$\pm$	&0.0064	&0.6240	&$\pm$	&0.0160	&3810	&$\pm$	&130	&	&	&\\
TrES-Her0-07621	&A	&0.4930	&$\pm$	&0.0030	&0.4530	&$\pm$	&0.0600	&3500	&$\pm$	&...  &1.137	&5	&\\
	&B	&0.4890	&$\pm$	&0.0030	&0.4520	&$\pm$	&0.0500	&3395	&$\pm$	& ...&	&	&\\
2MASS J05162881+2607387	&A	&0.7870	&$\pm$	&0.0120	&0.7880	&$\pm$	&0.0150	&4200	&$\pm$	&...&2.619	&6	&\\
	&B	&0.7700	&$\pm$	&0.0090	&0.8170	&$\pm$	&0.0100	&4154	&$\pm$	& ...&	&	&\\
UNSW-TR 2	&A	&0.5290	&$\pm$	&0.0350	&0.6410	&$\pm$	&0.0500	& ...&$\pm$	& ...	&2.144	&7	&\\
	&B	&0.5120	&$\pm$	&0.0350	&0.6080	&$\pm$	&0.0600	& ...	&$\pm$	& ...	&	&	&\\
NSVS 06507557	&A	&0.6560	&$\pm$	&0.0860	&0.6000	&$\pm$	&0.0300	&3960	&$\pm$	&80	&0.520	&8	&\\
	&B	&0.2790	&$\pm$	&0.0450	&0.4420	&$\pm$	&0.0240	&3365	&$\pm$	&80	&	&	&\\
NSVS 02502726	&A	&0.7140	&$\pm$	&0.0190	&0.6740	&$\pm$	&0.0600	&4300	&$\pm$	&200	&0.560	&9	&\\
	&B	&0.3470	&$\pm$	&0.0120	&0.7630	&$\pm$	&0.0050	&3620	&$\pm$	&205	&	&	&\\
T-Lyr1-17236	&A	&0.6795	&$\pm$	&0.0107	&0.6340	&$\pm$	&0.0430	&4150	&$\pm$	& ...	&8.430	&10	&\\
	&B	&0.5226	&$\pm$	&0.0061	&0.5250	&$\pm$	&0.0520	&3700	&$\pm$	& ... &	&	&\\
2MASS J01542930+0053266	&A	&0.6590	&$\pm$	&0.0310	&0.6390	&$\pm$	&0.0830	&3730	&$\pm$	&100	&2.639	&11	&\\
	&B	&0.6190	&$\pm$	&0.0280	&0.6100	&$\pm$	&0.0930	&3532	&$\pm$	&100	&	&	&\\
GJ 3236	&A	&0.3760	&$\pm$	&0.0170	&0.3828	&$\pm$	&0.0072	&3310	&$\pm$	&110	&0.770	&12	&\\
	&B	&0.2810	&$\pm$	&0.0150	&0.2992	&$\pm$	&0.0075	&3240	&$\pm$	&110	&	&	&\\
SDSS-MEB-1	&A	&0.2720	&$\pm$	&0.0200	&0.2680	&$\pm$	&0.0090	&3320	&$\pm$	&130	&0.410	&13	&\\
	&B	&0.2400	&$\pm$	&0.0220	&0.2480	&$\pm$	&0.0080	&3300	&$\pm$	&130	&	&	&\\
BD -22 5866	&A	&0.5881	&$\pm$	&0.0029	&0.6140	&$\pm$	&0.0450	& ...	&$\pm$	& ...	&2.211	&14	&\\
	&B	&0.5881	&$\pm$	&0.0029	&0.5980	&$\pm$	&0.0450	& ...	&$\pm$	& ...	&	&	&\\
NSVS 01031772	&A	&0.5428	&$\pm$	&0.0027	&0.5260	&$\pm$	&0.0028	&3615	&$\pm$	&72	&0.368	&15	&\\
	&B	&0.4982	&$\pm$	&0.0025	&0.5088	&$\pm$	&0.0030	&3513	&$\pm$	&31	&	&	&\\
NSVS 11868841	&A	&0.8700	&$\pm$	&0.0740	&0.9830	&$\pm$	&0.0300	&5260	&$\pm$	&110	&0.602	&8	&\\
	&B	&0.6070	&$\pm$	&0.0530	&0.9010	&$\pm$	&0.0260	&5020	&$\pm$	&110	&	&	&\\
GJ 2069A	&A	&0.4329	&$\pm$	&0.0018	&0.4900	&$\pm$	&0.0800	& ...	&$\pm$	& ...	&2.771	&16	&\\
	&B	&0.3975	&$\pm$	&0.0015	&0.3300	&$\pm$	&0.0400	& ...	&$\pm$	& ...	&	&	&\\
2MASS J04463285+1901432	&A	&0.4700	&$\pm$	&0.0500	&0.5700	&$\pm$	&0.0200	&3320	&$\pm$	&150	&0.630	&18	&\\
	&B	&0.1900	&$\pm$	&0.0200	&0.2100	&$\pm$	&0.0100	&2910	&$\pm$	&150	&	&	&\\
NSVS 6550671	&A	&0.5100	&$\pm$	&0.0200	&0.5500	&$\pm$	&0.0100	&3730	&$\pm$	&60	&0.193	&19	&\\
	&B	&0.2600	&$\pm$	&0.0200	&0.2900	&$\pm$	&0.0100	&3120	&$\pm$	&65	&	&	&\\
IM Vir	&A	&0.9810	&$\pm$	&0.0120	&1.0610	&$\pm$	&0.0160	&5570	&$\pm$	&100	&1.309	&20	&\\
	&B	&0.6644	&$\pm$	&0.0048	&0.6810	&$\pm$	&0.0130	&4250	&$\pm$	&130	&	&	&\\
RXJ0239.1	&A	&0.7300	&$\pm$	&0.0090	&0.7410	&$\pm$	&0.0040	&4645	&$\pm$	&20	&2.072	&21	&\\
	&B	&0.6930	&$\pm$	&0.0060	&0.7030	&$\pm$	&0.0020	&4275&$\pm$	&	15&	&	&\\
ASAS\,J045304-0700.4	&A	&0.8452	&$\pm$	&0.0056	&0.848	&$\pm$	&0.005	&5324	&$\pm$	&200	&1.6224	&22	&\\
	&B	&0.8390	&$\pm$	&0.0056	&0.833	&$\pm$	&0.005	&5105&$\pm$	&	200&	&	&\\
ASAS\,J082552-1622.8	&A	&0.703	&$\pm$	&0.003	&0.694	&$\pm$	&0.007	&4230	&$\pm$	&200	&1.52852	&22	&\\
	&B	&0.687	&$\pm$	&0.003	&0.699	&$\pm$	&0.011	&4280&$\pm$	&	200&	&	&\\	
ASAS\,J093814-0104.4	&A	&0.771	&$\pm$	&0.033	&0.772	&$\pm$	&0.012	&4360	&$\pm$	&200	&0.897442	&23	&\\
	&B	&0.768	&$\pm$	&0.021	&0.769	&$\pm$	&0.013	&4360&$\pm$	&	200&	&	&\\
ASAS\,J212954-5620.1	&A	&0.833	&$\pm$&0.017&0.845	&$\pm$	&0.012	&4750	&$\pm$	&150	&0.702430	&23	&\\
	&B	&0.703	&$\pm$	&0.013	&0.718	&$\pm$	&0.017	&4220 &$\pm$	&	180&	&	&\\

T-Cyg1-12664	&A	&0.680	&$\pm$&0.021&0.613	&$\pm$	&0.007	&4320	&$\pm$	&100	&4.12879779	&This study	&\\
	&B	&0.341	&$\pm$	&0.012	&0.897	&$\pm$	&0.012	&2750 &$\pm$	&	65&	&	&\\

\hline
\end{tabular}
\begin{list}{}{}
\item[Ref:]{ (1)\,\citet{table1}, (2)\,\citet{table2}, (3)\,\citet{table3}, (4)\,\citet{table4}, (5)\,\citet{table5}, (6)\,\citet{table6}, (7)\,\citet{table7},  
(8)\,\citet{table8}, (9)\,\citet{table9}, (10)\,\citet{2008AJ....135..850D}, (11)\,\citet{table11}, (12)\,\citet{table12}, (13)\,\citet{table13}, (14)\,\citet{table14}, 
(15)\,\citet{table15}, (16)\,\citet{table16}, (17)\,\citet{table17}, (18)\,\citet{table18}, (19)\,\citet{table19}  ,(20)\,\citet{table20}, (21)\,\citet{table21}, 
(22)\,\citet{table22}, (23)\,\citet{table23}}
\end{list}
\end{table*}

\section{Oversized Stars in the Low-Mass Eclipsing Binaries}
Abt (1963) and Duquennoy, Mayor and Halbwachs (1991) report that binary stars are more common 
than single stars at masses above that of the Sun. In contrary, Reid \& Gizis (1997) and Delfosse 
et al. (1999) suggest that binaries are not very common in low-mass stars, although approximately 
75 \% of all stars in our Galaxy are low-mass dwarfs with masses smaller than 0.7 \Msun. Due to the 
low binary fractions and their faintness very few low-mass eclipsing binary systems have been observed 
so far both photometrically and spectroscopically, yielding accurate physical parameters. 
Recently in a pioneer study, Ribas (2003, 2006) collected the available masses and radii of the low-mass stars
and compared with those obtained by stellar evolutionary models. Even small numbers of the sample, a total of eight 
double-lined eclipsing binaries whose masses and radii are determined with an accuracy of better than 3 \%, 
the comparison evidently revealed that the observed radii are systematically larger than the models. However, 
the effective temperatures are cooler than the theoretical calculations, being the luminosities in agreement 
with those of single stars with the same mass. This discrepancy between the models and observations has been explained 
by Mullan and MacDonald (2001), Torres et al. (2006), Ribas (2006) and Lopez-Morales (2007) and others by the high 
level magnetic activity in the low-mass stars. They suggest that stellar activity may be responsible for the observed 
discrepancy through inhibition of convection or effects of a significant spot coverage. However, Berger et al. (2006) 
compared the interferometric radii of low-mass stars with those estimated from theory and find a correlation between 
an increase in metallicity and larger-than-expected radius, i.e. the larger radii are caused by differences in metallicity. 

The number of double-lined low-mass eclipsing binaries having precise photometric and spectroscopic observations 
is now reached to 26. Combining the results of the analysis of both photometric and spectroscopic observations 
accurate masses, radii, effective temperatures and luminosities  of the components have been obtained. 
In Table\,7 we list absolute parameters for the low-mass stars with their standard deviations. 
In Fig.\,10 we show positions of the T-Cyg1-12664 components in the mass-radius (M-R) and mass - effective temperatures (M-T$_{\rm eff}$) planes relative to those of the well-determined 
low-mass stars in the eclipsing binary systems. Theoretical M-R diagrams for a zero-age main-sequence stars with [M/H]=0 
taken from the \citet{bar98} are also plotted for comparison.

\begin{figure}
\includegraphics[width=12cm]{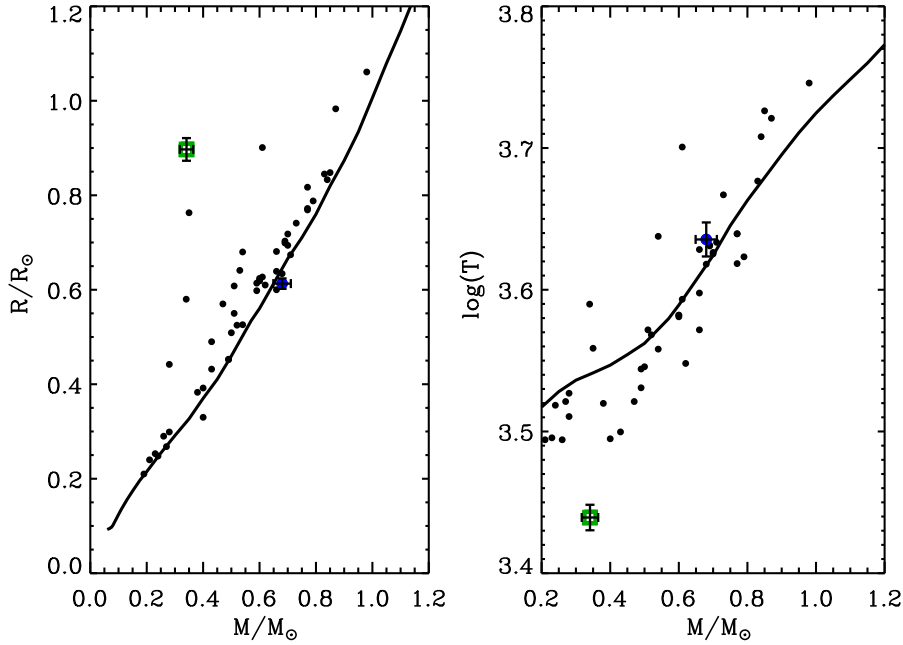}
\caption{Components of T-Cyg1-12664 (squares with error bars) in the mass-radius plane (left panel). The less massive  component is located among 
the most deviated stars from the theoretical mass-radius relationship \citep{cakirli}. The lines show stellar evolution models from  
\citet{bar98} for zero-age mian-sequence with [M/H]=0 (solid line). Components of T-Cyg1-12664 (squares with error bars) in the mass-effective temperature plane (right panel). The less massive  component appears to the coolest star with respect to the model. }  \end{figure}

A low-mass star has a convective envelope with a radiative core. The depth of the convective zone is about 
0.28 R for a star with a mass about 0.9 \Msun,~ and gradually increases to about 0.41 R for a 0.4 \Msun star.  Stars are 
thought to be fully convective below about 0.35 \Msun.~ Therefore, the internal structure of such a low-mass star with a deep 
convective zone is tightly dependent on the mixing length parameter, $\alpha$. \citet{cakirli} compared locations of the low-mass 
stars in the M-R and M-T$_{eff}$ diagrams with the theoretical calculations for various $\alpha$ parameters. Convection is modelled 
by mixing length theory (B{\"o}hm-Vitense 1981) with the ratio of mixing length to pressure scale height (i.e. $\alpha = l/H_p$). Theoretical 
models show that there is no significant separation in the  M-R relation for the stars below about 0.6 \Msun~, depending 
on the mixing length parameter. For masses above this value a separation is revealed. However, as the $\alpha$ parameter 
increases the computed effective temperature gets higher. A separation is clearly seen even for the masses below 0.6 \Msun~. Empirical 
parameters of the low-mass stars were also plotted in the same planes. This comparison clearly exposes that the discrepancies 
in radii and effective temperatures could not be explained only by changing $\alpha$ parameter as it is obvious in M-R and 
M-T$_{eff}$ relations. On the other hand Demory et al. (2009) showed that there is no significant correlation between metallicity
and radius of the single, low-mass stars. An alternative explanation remains to be the magnetic activity responsible for the 
observed larger radii but cooler effective temperatures. As demonstrated by Mullan \& MacDonald (2001) and Chabrier et al. (2007) magnetic fields change the evolution of low-mass stars. Due to the high 
magnetic activity in the fast-rotating dwarfs their surfaces are covered by dark spot(s) or spot groups. Spot coverage in active dwarfs 
yields larger radii and lower effective temperatures. The secondary component of T-Cyg1-12664 appears to be the mostly deviated star from the normal 
main-sequence stars in the M-R plane. The empirical relation between the mass and radius of low mass stars clearly shows that the stars 
with masses below 0.27~\Msun\, follow theoretical M-R relation. Thereafter a slight deviation begins from theoretical expectations and the 
largest deviation occurs for stars with a mass of about 0.34\,\Msun.\, Then, the deviations begin to decrease up to the solar mass. The
maximum deviation seen at a mass of 0.34\,\Msun\, which is very close to the mass of 0.35\,\Msun\, fully convective stars as suggested by theoretical studies \citep{reiners} .

\section{Discussion}
We obtained multi-band ,VRI light curves and spectra of T-Cyg1-12664. We 
analyzed the Kepler's R-data consisting of 5708 observations and radial velocities simultaneously. The light contribution of the secondary star $L_2$/($L_1$+$L_2$)=0.031 and 0.102 were obtained for the V-, and R-bandpass, respectively. This result indicates that the light contribution of the less massive component is very small, indicating its effect on the color at out-of-eclipse is very limited for the shorter wavelengths. Kepler's R-data clearly show that there is a wave-like distortion on the light curve. Both the amplitude and the shape of these distortions vary with time. These distortions are modelled by two separate spots on the primary star, indicating two active longitude belts. In addition a sine-like variation in the mean normalized light of the system with a peak-to-peak amplitude of 0.007  
in light units has been revealed. Extracting light contribution of  the secondary star  to total light and adopting various bolometric contributions for a K5V star we estimated a distance of  127$\pm$14 pc for the system. 

Comparison of the theoretical models shows that the secondary star has a radius of 2.8 times larger than that expected for its mass. It is located amongst the 
largest deviated low-mass stars from the mass-radius relation of the zero-age main-sequence stars. However, its effective temperature seems to be the lowest with respect to the models for solar composition. The mass of the secondary star is obtained with an accuracy of about 3.5 \% which is very close to the mass of 0.35\,\Msun\, which represents stars passing from partially convective envelope to the fully convective ones.

\section*{Acknowledgments}
We thank to T\"{U}B{\.I}TAK National Observatory (TUG) for a partial support in using RTT150 and T100 telescopes with project numbers 
10ARTT150-483-0, 11ARTT150-123-0 and 10CT100-101. 
We also thank to the staff of the Bak{\i}rl{\i}tepe observing station for their warm hospitality. 
We thank the referee for a timely and useful report.
The following internet-based resources were used in research for this paper: the NASA Astrophysics Data System; the SIMBAD database operated at CDS, Strasbourg, France; T\"{U}B\.{I}TAK 
ULAKB{\.I}M S\"{u}reli Yay{\i}nlar Katalo\v{g}u-TURKEY; and the ar$\chi$iv scientific paper preprint service operated by Cornell University. We are grateful to the anonymous referee, whose comments and suggestions helped to improve this paper. 

\bibliographystyle{mn_new}

\label{lastpage}

\end{document}